\documentclass[onecolumn,conference,letterpaper]{IEEEtran}

\IEEEoverridecommandlockouts                              

\usepackage[utf8]{inputenc} 
\usepackage[T1]{fontenc}
\usepackage{url}
\usepackage{ifthen}
\usepackage{cite}
\usepackage[cmex10]{amsmath} 

\usepackage{color}
\usepackage{graphicx}
\usepackage{amssymb,mathtools}
\usepackage{bm}
\usepackage{comment}
\usepackage[colorlinks=true]{hyperref}

\usepackage{algorithm,algorithmic}

\DeclareMathOperator{\Hessian}{Hess}

\hypersetup{
     colorlinks = true,
     linkcolor = [rgb]{0,0,1},
     anchorcolor = [rgb]{0,0,1},
     citecolor = [rgb]{0.9,0.5,0},
     filecolor = [rgb]{0,0,1},
     urlcolor = [rgb]{0.9,0.5,0},
        bookmarksopen = true,
        bookmarksnumbered = true,
        breaklinks = true,
        colorlinks = true,
        linkcolor = [rgb]{0.2,0.6,0.2},
        urlcolor  = [rgb]{0,0,1},
        citecolor = [rgb]{0.9,0.5,0},
        anchorcolor = [rgb]{0.2,0.6,0.2},
}

\usepackage{todonotes}
\usepackage{lipsum} 
\usepackage{subcaption}
\newtheorem{thm}{\bf{Theorem}}[section]
\newtheorem{lemma}{\bf{Lemma}}[section]
\newtheorem{assumption}{\bf{Assumption}}[section]

\newtheorem{remark}{\bf{Remark}}[section]

\newcommand{\nn}{\nonumber}

\title{\LARGE \bf
Privacy Signaling Games with Binary Alphabets}

\author{Photios~A.~Stavrou$^{1}$, Serkan~Sar{\i}ta\c{s}$^{2}$ and Mikael Skoglund$^{3}$
\thanks{*This work was funded in part
by the Swedish research council under contract 2019-03606.}
\thanks{{$^{1}$Photios A. Stavrou is with the Communication Systems Department at EURECOM, Campus SophiaTech, 06904, France. {\tt\small fotios.stavrou@eurecom.fr}}}
\thanks{$^{2}$Serkan~Sar{\i}ta\c{s} is with the Department of Electrical and Electronics Engineering, Middle East Technical University, 06800, Ankara, Turkey.
        {\tt\small ssaritas@metu.edu.tr}}%
\thanks{{$^{3}$ Mikael Skoglund is with the Division of Information Science and Engineering, KTH Royal Institute of Technology, SE-10044, Stockholm, Sweden. {\tt\small skoglund@kth.se}}}
}

\begin{document}

\maketitle
\thispagestyle{empty}
\pagestyle{empty}

\begin{abstract}
In this paper, we consider a privacy signaling game problem for binary alphabets and single-bit transmission where a transmitter has a pair of messages, one of which is a casual message that needs to be conveyed, whereas the other message contains sensitive data and needs to be protected. The receiver wishes to estimate both messages to acquire as much information as possible. For this setup, we study the interactions between the transmitter and the receiver with non-aligned information-theoretic objectives (modeled by mutual information and hamming distance) due to the privacy concerns of the transmitter. We derive conditions under which Nash and/or Stackelberg equilibria exist and identify the optimal responses of the encoder and decoders strategies for each type of game. One particularly surprising result is that when both types of equilibria exist, they admit the same encoding and decoding strategies. We corroborate our analysis with simulation studies. 

\end{abstract}

\section{Introduction}\label{sec:intro}

\par Decision-making is pivotal for a wide range of real-world applications such as social networks, networked control systems, smart grids, and recommendation systems. In these applications, usually, several customers (users) in a network may share extensive {amounts} of information with some service provider (i.e., utility company) because the latter wishes to know as much as possible about the service offered at the customer to improve the quality of service. However, this may come with a price as sometimes the users in the network may be prone to network-based attacks from malicious elements aiming to steal some sensitive information. Therefore, the users, in addition to the continuous improvement of the quality of service offered by a provider, wish to maintain a certain level of privacy. A type of privacy objective can be assumed when the information transmitted by some user to the service provider may be correlated with certain private information they want to protect. For example, in smart grids, the smart meter provides real-time information on energy supplies from the energy provider on the demands of the consumer (user), which can be utilized for unauthorized purposes, e.g., to infer the private information of the consumer, such as their habits and behaviors, see, e.g., \cite{mcdaniel-mclaughlin:2009,finster-baumgart:2015}. Identifying privacy-preserving mechanisms or approaches under various contexts related to information theory and control applications can be found in an anthology of papers, for instance, in \cite{ny-2014,gomez-vilardebo:2015,zuxing:2019,nekouei:2019,lu-zhu:2020,cavarec:2021}.

\subsection{Motivational Example}\label{subsec:mot_example}

\par
\begin{figure} [htp]
	\begin{center}
\includegraphics[width=0.6\columnwidth]{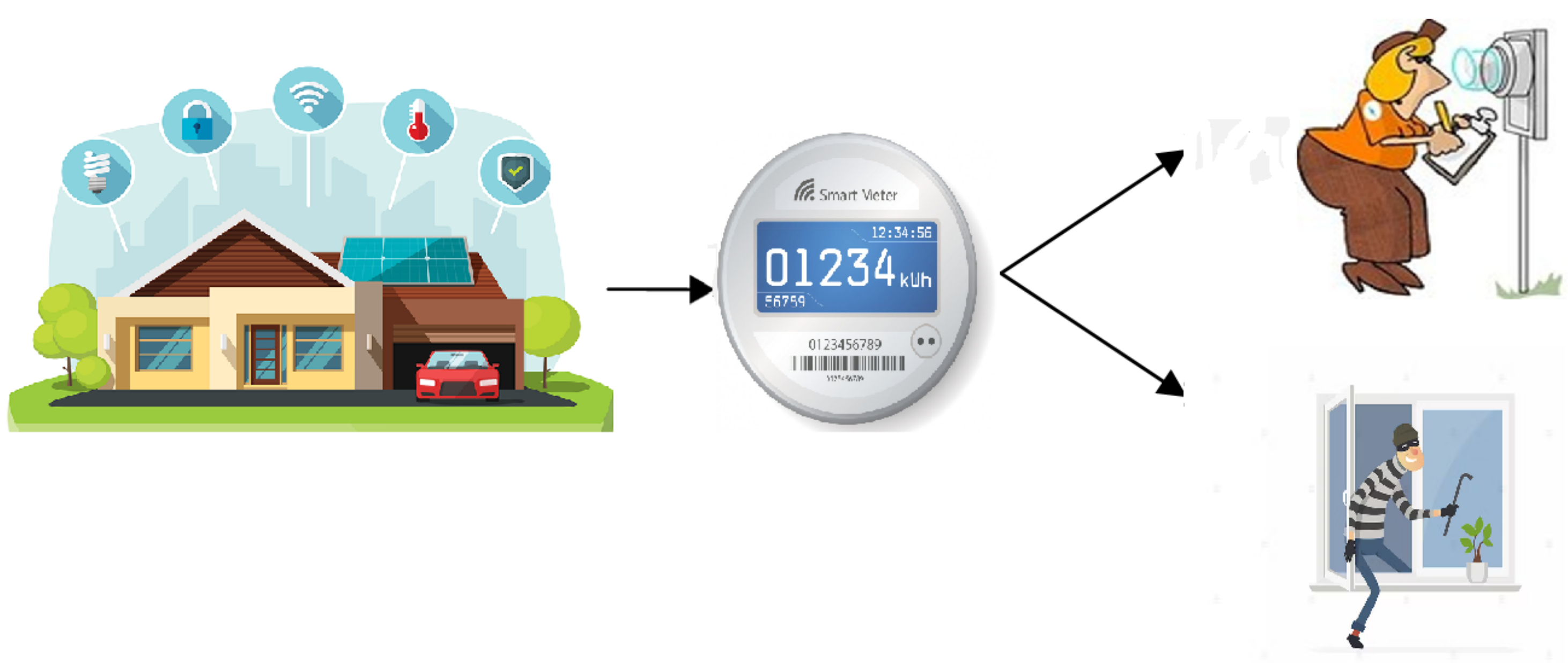}
	\end{center}
	\caption{Motivational example.}
	\label{fig:privacy_signaling_games_motivation}
\end{figure}

 Consider the scenario illustrated in Fig.\ref{fig:privacy_signaling_games_motivation}. In that scenario, a smart house is illustrated in which a smart meter records the energy consumption and exchanges consumption data with energy suppliers, which can be used for monitoring and billing. Evidently, the existence of home residents and the electricity usage recorded by the smart meter at home are, in practice, correlated. Nevertheless, the presence of the house residents at home should be kept secret to possible outsiders (burglars, adversaries, etc.) whereas, at the same time, the energy consumption should be available to the electricity service providers. Therefore, the smart meter should be designed so that the service providers can access the electricity usage data, whereas the outsiders should not be able to deduce if the residents are home or not by checking the information from the smart meter. 

\subsection{Literature review}\label{subsec:lit_review}

\par The studies on cheap talk and signaling games were initiated by Crawford and Sobel in their seminal work\cite{SignalingGames}, and found applications in various topics, e.g., in networked systems \cite{misBehavingAgents,csLloydMax}, recommendation systems \cite{miklos2013value,recommSystemGame}, and economics \cite{signalSurvey,Sobel2009}. Starting with \cite{bayesianPersuasion}, there are many studies that consider the Stackelberg equilibrium of signaling games; an incomplete list includes \cite{tacWorkSerkan,CedricWork,akyolITapproachGame,omerHierarchial,dynamicGameSerkan,strategicCommSideInfo,persuasionLimCap, serkanACC2020,ertanMismatch,saritas-stavrou:2021} (see also the references therein). Many of these works assume that the non-alignment between the objective functions of the encoder and the decoder is a function of a Gaussian random variable (RV) correlated with the Gaussian source and hidden from the decoder (unlike the original case where it is fixed and commonly known by the encoder and the decoder \cite{SignalingGames}, which is also studied in \cite{tacWorkSerkan,dynamicGameSerkan,serkanACC2020}). Nash and Stackelberg equilibria of signaling games are investigated in \cite{ertanMismatch} when there is a mismatch in priors of players. We refer to \cite{Sobel2009,tacWorkSerkan,dynamicGameSerkan} for more discussion on the literature and some extensions (including Nash equilibrium analyses and multi-stage extensions) on cheap talk and signaling games. 

In the context of strategic information transmission, several works consider the scenario where the sender takes the privacy of certain information into account by deploying a suitable privacy measure under either the Nash or Stackelberg criteria. For instance, in \cite{farokhi:2015}, a communication scenario between a sender and a receiver is investigated using the Stackelberg equilibrium. A family of nontrivial equilibria, in which
the communicated messages carry information, is constructed,
and its properties are studied. In \cite{akyol:2015}, the authors study a Stackelberg game where the utility measure for
the public parameter is quadratic and the privacy measure
is entropy-based. Additional results therein include characterizations of the equilibrium under noisy and noiseless communication scenarios and analysis of the corresponding coding policies. In \cite{farokhi:2016}, the effect of privacy via a Nash game is studied between a sender and a receiver. As a measure of merit, the authors use mutual information between the private information and the communicated message to quantify the amount of the leaked information. For discrete RVs, they provide a numerical algorithm to find an equilibrium, whereas for Gaussian RVs, a bound on the estimation error is provided, and affine policies are shown to achieve this bound.
In \cite{kazikli:2020}, a privacy-signaling game problem for the setup in Fig. \ref{fig:privacy_signaling_games} is considered in which a transmitter with privacy concerns observes a pair of correlated random vectors which are modeled as jointly Gaussian. Among other results, it is shown that a payoff dominant Nash equilibrium among all admissible policies is attained by a set of explicitly characterized linear policies and coincides with a Stackelberg equilibrium. {Regarding the state of the art privacy metrics, we refer to \cite{privacyMetricSurvey} for a selection of over eighty privacy metrics and their categorization. Herein, we consider one of the discussed metrics therein, namely, Hamming distance.}

In this paper, we consider the scenario that was first introduced in \cite{kazikli:2020}. In this setup, a transmitter has {a pair of messages, one of which} is a casual message that needs to be conveyed, whereas the other message contains sensitive data and needs to be protected. On the other hand, the receiver wishes to estimate both messages with the aim of acquiring as much information as possible. For this setup, we study the interactions between the transmitter and the receiver whose objectives are not-aligned due to the privacy concerns of the transmitter in a game-theoretic framework. However, in contrast to \cite{kazikli:2020} that deals with jointly Gaussian random vectors {(of possibly different lengths)} and linear policies, {\it here we deal with {binary alphabets} and consider different objectives for our single-bit transmitter and receiver}. 

\subsection{Contributions}\label{subsec:contributions}

The main contributions of this paper can be summarized as follows:
\begin{itemize}
    \item[(i)] We model a binary privacy signaling game, {assuming a single-bit transmission}, between an encoder and a decoder in which the encoder aims to hide one of two correlated binary RVs (i.e., private message) and to transmit the other (i.e., public message) while the decoder's goal is to learn about both of the RVs as much as possible. We use mutual information as a metric to measure the information exchange, a Hamming distortion to measure the level of privacy, and a weighting coefficient that determines the importance of privacy from the perspective of the encoder.
    \item[(ii)] We characterize the objective functions of the encoder and the decoder in terms of priors and strategies (see Lemma~\ref{lemma:characterization}), derive the best response of the encoder (resp. decoder) for a given decoder (resp. encoder) (see Lemmas~\ref{lemma:bestResponseEncoder} and \ref{lemma:bestResponseDecoder}). Then, using these best response maps, we characterize the Stackelberg and Nash equilibria (see Theorems~\ref{thm:stackelberg} and \ref{thm:nash}, respectively).
    \item[(iii)] We show that under certain conditions on the source prior probabilities, Nash and Stackelberg equilibria exist and coincide. Otherwise, there does not exist a Nash equilibrium, and the privacy coefficient only affects the Stackelberg equilibrium. In particular, for large privacy coefficients, the encoder may even prefer to hide information about the public message.
\end{itemize}

\section{Problem formulation and preliminaries}\label{sec:prob_form}

\begin{figure} [htp]
	\begin{center}
\includegraphics[width=0.6\columnwidth]{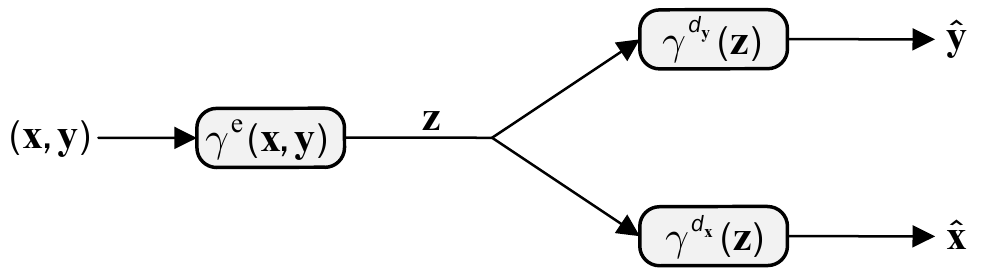}
	\end{center}
	\caption{Our setup.}
	\label{fig:privacy_signaling_games}
\end{figure}

In this paper, we consider the scenario illustrated in Fig.~\ref{fig:privacy_signaling_games} that was first introduced in \cite{kazikli:2020}.
We assume that the transmitter encodes a pair of correlated random variables $({\bf x}, {\bf y})\in{\cal X}\times{\cal Y},\; {\cal X}={\cal Y}=\{0,1\}$ into ${\bf z}\in{\cal Z}=\{0,1\}$ using an encoding function denoted by ${z}=\gamma^{e}(x,y)$ and the receiver wants to decode both messages based on the observation ${\bf  z} = z$. Note that the transmitter desires to transmit information about ${\bf y}$ and sees ${\bf x}$ as a private parameter that needs to {be hidden} from the receiver. In contrast, the receiver wants to accurately estimate both public and private messages given the observation ${\bf z}= z$. We denote the decoding functions for estimating ${\bf x}$ and ${\bf y}$ by $\hat{x}=\gamma^{d_{\bf x}}({z})$ and $\hat{y}=\gamma^{d_{\bf y}}({z})$, respectively.

\par Since the transmitter needs to encode two messages generated by the joint distribution of ({\bf x}, {\bf y}), i.e., ${\bf P}(x,y)$, it means that hiding ${\bf x}$ or transmitting ${\bf y}$ are somehow inter-dependent actions. Since our scenario is for binary alphabets, in the sequel, we will denote the joint distribution or probability mass function of $({\bf x}, {\bf y})$ to be given by the following column stochastic matrix:
\begin{align}
{\bf P}(x,y)=\begin{bmatrix}{\bf P}(x=0,y=0)\\
{\bf P}(x=0,y=1)\\
{\bf P}(x=1,y=0)\\
{\bf P}(x=1,y=1)
\end{bmatrix}=\begin{bmatrix}a\\
b\\
c\\
d
\end{bmatrix}, 
\label{joint_prob_input_messages}    
\end{align}
where $d\triangleq{1-(a+b+c)}$ with $(a, b, c, d)\in[0,1]\times[0,1]\times[0,1]\times[0,1]$ (also denoted for simplicity $[0,1]^4$).
The objective of the transmitter is to maximize the public information ${\bf y}$ for the receiver and at the same time to hide as much as possible the sensitive information ${\bf x}$. These can be cast by the following objective function

\begin{align}
J^e(\gamma^e,\gamma^{d_{\bf x}},\gamma^{d_{\bf y}})=I({\bf y};\hat{\bf y})+\rho{\bf E}\{d_H({\bf x},\hat{\bf  x})\},\label{obj_fun_enc}    
\end{align}
which is to be maximized by the encoder, where $I({\bf y};\hat{\bf y})$ is the mutual information between ${\bf y}$ and $\hat{\bf y}$ \cite{cover-thomas:2006}, $\rho>0$ is a weighting coefficient that determines the level of desired privacy of ${\bf x}$, 
${\bf E}\{d_H({\bf x},\hat{\bf x})\}$ is some loss function which for this paper is assumed to be modeled by Hamming distortion, i.e., 
\begin{align}
    d_H&({\bf x},\hat{\bf  x}) = \begin{cases} 
      1 & {\bf x} \neq \hat{\bf  x} \\
      0 & {\bf x} = \hat{\bf  x}
   \end{cases}, \label{distortion_function}
\end{align}   
responsible to capture the privacy term ${\bf x}$.
The objective of the receiver is to maximize the information of both public information ${\bf y}$  and sensitive information ${\bf x}$. This can be cast by the following objective function
\begin{align}
J^d(\gamma^e,\gamma^{d_{\bf x}},\gamma^{d_{\bf y}})=I({\bf y};\hat{\bf y})-{\bf E}\{d_H({\bf x},\hat{\bf  x})\}\,,\label{obj_fun_dec}    
\end{align}
which is to be maximized by the decoder. Since the costs of the encoder and the decoder are not aligned, the problem is studied under a game theoretic framework, and Stackelberg and Nash equilibria are investigated. In the Nash (simultaneous-move) game, the encoder and the decoder announce their strategies at the same time. {More precisely, suppose that the set of possible strategies at the encoder is denoted by $\Gamma^e$ and those at the decoders by $\Gamma^{d_{\bf y}}$ and $\Gamma^{d_{\bf x}}$, respectively, such that $\gamma^e\in\Gamma^{e}$, $\gamma^{d_{\bf y}}\in\Gamma^{d_{\bf y}}$, $\gamma^{d_{\bf x}}\in\Gamma^{d_{\bf x}}$}. {Then}, a triplet of policies $(\gamma^{e,*}, \gamma^{d_{\bf y},*}, \gamma^{d_{\bf x},*})$ is said to be a {\it Nash equilibrium} \cite{basols99} if
\begin{align}
\begin{split}
J^e(\gamma^{e,*},\gamma^{d_{\bf y},*}, \gamma^{d_{\bf x},*}) &\geq J^e(\gamma^{e}, \gamma^{d_{\bf y},*}, \gamma^{d_{\bf x},*}), \quad \forall \gamma^e \in \Gamma^e \,,\\
J^d(\gamma^{e,*}, \gamma^{d_{\bf y},*}, \gamma^{d_{\bf x},*})
&\geq J^d(\gamma^{e,*}, \gamma^{d_{\bf y}}, \gamma^{d_{\bf x}}) \quad \forall \gamma^{d_{\bf y}} \in \Gamma^{d_{\bf y}}, \gamma^{d_{\bf x}} \in \Gamma^{d_{\bf x}} \,.
\end{split}\label{eq:nashEquilibrium}
\end{align}
As observed in \eqref{eq:nashEquilibrium}, none of the players prefer to change their optimal strategies at the equilibrium, i.e., there is no profitable unilateral deviation from any of the players. In the Stackelberg game, the leader (encoder) commits to a particular policy and announces it to the follower (decoder). The decoder takes its optimal action upon observing the encoder's committed strategy. More precisely, a triplet of strategies $(\gamma^{e,*}, \gamma^{d_{\bf y},*}, \gamma^{d_{\bf x},*})$  is said to be a {\it Stackelberg equilibrium} \cite{basols99} if
\begin{align}
\begin{split}
J^e(\gamma^{e,*}, \gamma^{d_{\bf y},*}(\gamma^{e,*}), \gamma^{d_{\bf x},*}(\gamma^{e,*})) &\geq J^e(\gamma^e, \gamma^{d_{\bf y},*}(\gamma^e), \gamma^{d_{\bf x},*}(\gamma^e)), \quad \forall \gamma^e \in \Gamma^e \,,\\
\text{where } &(\gamma^{d_{\bf y},*}(\gamma^e), \gamma^{d_{\bf x},*}(\gamma^e)) \text{ satisfy} \\
J^d(\gamma^{e}, \gamma^{d_{\bf y},*}(\gamma^{e}),  \gamma^{d_{\bf x},*}(\gamma^{e})) &\geq J^d(\gamma^{e}, \gamma^{d_{\bf y}}(\gamma^{e}), \gamma^{d_{\bf x}}(\gamma^{e})) \quad \forall \gamma^{d_{\bf y}} \in \Gamma^{d_{\bf y}}, \gamma^{d_{\bf x}} \in \Gamma^{d_{\bf x}}  \,.\nn
\end{split}
\end{align}
Note that the follower (decoder) takes its action after observing the strategy $\gamma^{e}$ of the leader (encoder), thus the strategies $(\gamma^{d_{\bf y}}(\gamma^{e}), \gamma^{d_{\bf x}}(\gamma^{e}))$ of the decoder are a function of $\gamma^{e}$. 

\section{Main Results}\label{sec:main_results}

\par Before we start with our main results, we first
introduce the general structure of the ``stochastic'' encoder and decoder policies for our setup. {In particular, the encoder is given by the transition matrix
\begin{align}
{\bf P}^{e}(z|x,y)=\begin{bmatrix}
\kappa_1 & \kappa_2 & \kappa_3 & \kappa_4 \\
1-\kappa_1 & 1-\kappa_2 & 1-\kappa_3 & 1-\kappa_4
\end{bmatrix},\label{trans_prob_enc}
\end{align}
where $(\kappa_1, \kappa_2, \kappa_3, \kappa_4)\in[0,1]^4$}, whereas the transition matrices at the decoder are given by the column stochastic matrices
\begin{align}
&{\bf P}^{d_{\bf y}}(\hat{y}|z)=\begin{bmatrix}
\delta_1 & \delta_2  \\
1-\delta_1 & 1-\delta_2
\end{bmatrix},\label{trans_prob_dec_1}\\
&{\bf P}^{d_{\bf x}}(\hat{x}|z)=\begin{bmatrix}
\epsilon_1 & \epsilon_2  \\
1-\epsilon_1 & 1-\epsilon_2
\end{bmatrix},\label{trans_prob_dec_2}
\end{align}
where $(\delta_1, \delta_2, \epsilon_1, \epsilon_2)\in[0,1]^4$.
To derive our main results, we make use of the following assumption.
\begin{assumption}\label{assumption:1}(Structural assumption on \eqref{trans_prob_enc}) Restrict the information structure in \eqref{trans_prob_enc} to one that $\kappa_3=1-\kappa_2$ and $\kappa_4=1-\kappa_1$.
\end{assumption}

\begin{remark}\label{rem:commOnAssumption}(Comments on Assumption~\ref{assumption:1})
By putting such a restriction (i.e., a ``symmetric'' encoder assumption) on $\kappa_3$ and $\kappa_4$, we prevent infinitely many quadruples $(\kappa_1,\kappa_2,\kappa_3,\kappa_4)$ resulting in essentially equivalent encoders with respect to performance. Furthermore, after eliminating redundant quadruples by Assumption~\ref{assumption:1}, it is possible to obtain the (joint) convexity of $I({\bf y};\hat{\bf y})$ with respect to $\kappa_1$ and $\kappa_2$ in Appendix~\ref{proof:lemma:bestResponseEncoder}. Otherwise, i.e., without Assumption~\ref{assumption:1}, there is no conclusion on the (joint) convexity/concavity of $I({\bf y};\hat{\bf y})$ with respect to the quadruple $(\kappa_1,\kappa_2,\kappa_3,\kappa_4)$. 
\end{remark}
Next, we prove a lemma that reformulates the objective functions of \eqref{obj_fun_enc}, \eqref{obj_fun_dec}. We note that this lemma holds even if Assumption \ref{assumption:1} does not hold.
\begin{lemma}\label{lemma:characterization}(Characterization) For the information structure of the stochastic encoder and decoder in \eqref{trans_prob_enc}-\eqref{trans_prob_dec_2}, the objective functions in \eqref{obj_fun_enc}, \eqref{obj_fun_dec} can be characterized as follows
\begin{align}
&J^e(\gamma^e,\gamma^{d_{\bf x}},\gamma^{d_{\bf y}})=I({\bf y};\hat{\bf y})+\rho{\bf E}\{d_H({\bf x},\hat{\bf  x})\},\label{reform_obj_enc}\\ 
&J^d(\gamma^e,\gamma^{d_{\bf x}},\gamma^{d_{\bf y}})=I({\bf y};\hat{\bf y})-{\bf E}\{d_H({\bf x},\hat{\bf  x})\},\label{reform_obj_dec}
\end{align}
where\footnote{The logarithms are taken with base two throughout the paper.}
\begin{align}
  &I({\bf y};\hat{\bf y}) = H_b (q_1) + H_b (P_1 + P_2) + P_1 \log(P_1) + P_2\log(P_2)+  (q_1-P_1)\log(q_1-P_1)+ (1-q_1-P_2)\log(1-q_1-P_2) \,,\label{char_mutual_info}\\
  &{\bf E}\{d_H({\bf x},\hat{\bf  x})\} = a (1-n_1) + b(1-n_2) + cn_3 + dn_4,  \label{eq:hammingLoss}
\end{align}
with $H_b(p)$ denoting the binary entropy function, i.e.,  $H_b(p)\triangleq -p \log p - (1-p)\log (1-p)$, 
$q_1 \triangleq a+c$, $P_1 \triangleq a t_1 + ct_3$, $P_2 \triangleq b t_2 + dt_4$, $t_i \triangleq \delta_1\kappa_i + \delta_2(1-\kappa_i)$ and $n_i \triangleq \epsilon_1\kappa_i + \epsilon_2(1-\kappa_i)$ for $i=1,2,3,4$.
\end{lemma}
\IEEEproof See Appendix \ref{proof:lemma:characterization}.
\endIEEEproof

After formulating the objectives of the encoder and the decoder, next, we characterize their optimal strategies, in particular, their best responses for any other given strategy.

\begin{lemma}\label{lemma:bestResponseEncoder}(Best Response: Encoder)
Suppose that Assumption \ref{assumption:1} holds. Then, for given decoder strategies $\gamma^{d_{\bf x}}$ and $\gamma^{d_{\bf y}}$, the objective function of the encoder in \eqref{reform_obj_enc} is a jointly convex function of the pair $(\kappa_1, \kappa_2)$, and the maximum is achieved at one of the extreme points, i.e., $\kappa_1\kappa_2=\{00,01,10,11\}$.
\end{lemma}
\IEEEproof See Appendix \ref{proof:lemma:bestResponseEncoder}.
\endIEEEproof

\begin{lemma}\label{lemma:bestResponseDecoder}(Best Response: Decoder) Suppose that Assumption \ref{assumption:1} holds. Then the following hold.
\begin{itemize}
    \item[(i)] For a given encoder strategy $\gamma^{e}$, $I({\bf y};\hat{\bf y})$ in \eqref{char_mutual_info} is a jointly convex function of the pair $(\delta_1, \delta_2)$, and the maximum is achieved either when $\delta_1\delta_2=01$ or $\delta_1\delta_2=10$.
    \item[(ii)] For a given encoder strategy $\gamma^{e}$, the average distortion ${\bf E}\{d_H({\bf x},\hat{\bf  x})\}$ in \eqref{eq:hammingLoss} is minimized using the decoder strategy characterized in Table~\ref{table:optDecoderX}, where $\theta\triangleq \kappa_1(a+d)+\kappa_2(b+c)$.  
\begin{table}[ht]
\caption{Optimal decoder strategy to minimize the average distortion.}
\label{table:optDecoderX}
\centering
\begin{tabular}{|c|c|c|c|}
\hline
Condition & $\epsilon_1$ & $\epsilon_2$ & ${\bf E}\{d_H({\bf x},\hat{\bf  x})\}$  \\ \hline
$a+b \leq \theta \leq c+d$ & $0$ & $0$ & $a+b$  \\ \hline
$\theta \leq a+b \,, \, \theta \leq c+d$ & $0$ & $1$ & $\theta$  \\ \hline
$\theta \geq a+b \,, \, \theta \geq c+d$ & $1$ & $0$ & $1-\theta$  \\ \hline
$a+b \geq \theta \geq c+d$ & $1$ & $1$ & $c+d$  \\ \hline
\end{tabular}
\end{table}
\end{itemize}

\end{lemma}
\IEEEproof See Appendix \ref{proof:lemma:bestResponseDecoder}.
\endIEEEproof
Next, we proceed to derive conditions for which Nash and/or Stackelberg equilibria exist together with their corresponding optimal strategies.
\begin{thm}\label{thm:stackelberg}(Stackelberg)
Suppose that Assumption \ref{assumption:1} holds and $\min\{a+b, c+d, a+d, b+c\}$ is $a+b$ or $c+d$. Then, the following strategies form a Stackelberg equilibrium:
    \begin{align}
    \begin{split}
    \kappa_1\kappa_2 &= 01 \text{ or } \kappa_1\kappa_2 = 10  ~~\text{(Encoder)}\\
    \delta_1\delta_2 &= 01 \text{ or } \delta_1\delta_2 = 10 ~~\text{(Decoder-$\hat{\bf y}$)}\\
    \epsilon_1\epsilon_2 &= \begin{cases}
    00 \text{ if } a+b \leq c+d \\
    11 \text{ if } a+b \geq c+d
    \end{cases}.~~\text{(Decoder-$\hat{\bf x}$)}
    \end{split}\label{stack:strategies}
    \end{align}
    
    Otherwise, if $\min\{a+b, c+d, a+d, b+c\}$ is $a+d$ or $b+c$, then, for sufficiently small $\rho$, the equilibrium strategies of \eqref{stack:strategies} are still valid. In contrast, for sufficiently large $\rho$, the decoder strategies in \eqref{stack:strategies} are still the same and the optimum encoder strategy lies at the boundary of the $\kappa_1\kappa_2$ region which satisfy either $a+b \leq \theta \leq c+d$ or $a+b \geq \theta \geq c+d$, where $\theta\triangleq \kappa_1(a+d)+\kappa_2(b+c)$ (defined as before). 
\end{thm}
\IEEEproof See Appendix \ref{proof:thm:stackelberg}.
\endIEEEproof

\begin{thm}\label{thm:nash}(Nash)
Suppose that Assumption \ref{assumption:1} holds and $\min\{a+b, c+d, a+d, b+c\}$ is $a+b$ or $c+d$. Then, the same strategies as in \eqref{stack:strategies} form a Nash equilibrium. Otherwise, there does not exist a Nash equilibrium.\footnote{We exclude the trivial case of equal priors $a=b=c=d=0.25$ in which any strategy pair ends up an equilibrium.}
\end{thm}
\IEEEproof See Appendix \ref{proof:thm:nash}.
\endIEEEproof
Next we give some technical comments related to our results in Theorems \ref{thm:stackelberg}, \ref{thm:nash}.
\begin{remark}\label{rem:techComm}(Technical comments)
{\bf (TC1)} When in Theorems \ref{thm:stackelberg}, \ref{thm:nash}, $\min\{a+b, c+d, a+d, b+c\}$ is $a+b$ or $c+d$, the optimal encoder selects either $\kappa_1\kappa_2=01$ or $\kappa_1\kappa_2=10$, which correspond to sending information only about ${\bf y}$ (e.g.,  $\kappa_1\kappa_2=01$ implies ${\bf P}^e(z|x,y)={\bf P}^e(z|y)$). In this case, since the received message ${\bf z}$ does not contain any direct information about ${\bf x}$, the decoder-$\hat{\bf x}$ uses only priors of ${\bf P}(x)$ and achieves the average distortion ${\bf E}\{d_H({\bf x},\hat{\bf  x})\}=\min\{a+b, c+d\}$. {\bf (TC2)} If $\min\{a+b, c+d, a+d, b+c\}$ is $a+d$ or $b+c$ and the encoder still uses $\kappa_1\kappa_2=01$ or $\kappa_1\kappa_2=10$, then, the decoder makes use of the conditional probability ${\bf P}(x|y)$ (since ${\bf z}$ is directly related to ${\bf y}$), which further means that the average distortion ${\bf E}\{d_H({\bf x},\hat{\bf  x})\}=\min\{a+d, b+c\}$. Hence in order to increase the privacy level (i.e., increase the average distortion to $\min\{a+b, c+d\}$), the encoder uses different strategies that result in smaller value of mutual information (see Fig.~\ref{fig:stackelbergExample}). To make this point clear, we note that the strategies in \eqref{stack:strategies} result in $I({\bf y};\hat{\bf y}) =H_b (q_1)$ and ${\bf E}\{d_H({\bf x},\hat{\bf  x})\} = \min\{a+b, c+d\}$. However, for large privacy weighting coefficient $\rho$, as shown in Theorem~\ref{thm:stackelberg}, the average distortion does not change, and the mutual information, as a function of the pair $(\kappa_1,\kappa_2)$, can be characterized as in \eqref{char_mutual_info} with $P_1=a\kappa_1+c(1-\kappa_2)$ and $P_2=b\kappa_2+d(1-\kappa_1)$. The resulting mutual information value will be less than $H_b(q_1)$, which means that the encoder ventures to send less information about the public message to be able to hide information about the private message. 
\end{remark}

\begin{remark}(Connection to similar work)
In \cite{kazikli:2020}, a similar setup is considered in which the random sources are jointly Gaussian, and the squared error is utilized as a privacy and information exchange metric. Similar to our result, it is shown that Stackelberg and payoff dominant Nash equilibria coincide. However, due to the difference between our source assumption (i.e., binary), information exchange metric (i.e., mutual information), and privacy metric (i.e., Hamming distortion), we have some cases under which Stackelberg equilibria exist, but there is no Nash equilibrium. 
\end{remark}

\section{Numerical Results}

\begin{figure}
	\centering
	\begin{subfigure}[t]{0.30\textwidth}
		\includegraphics[scale=0.35]{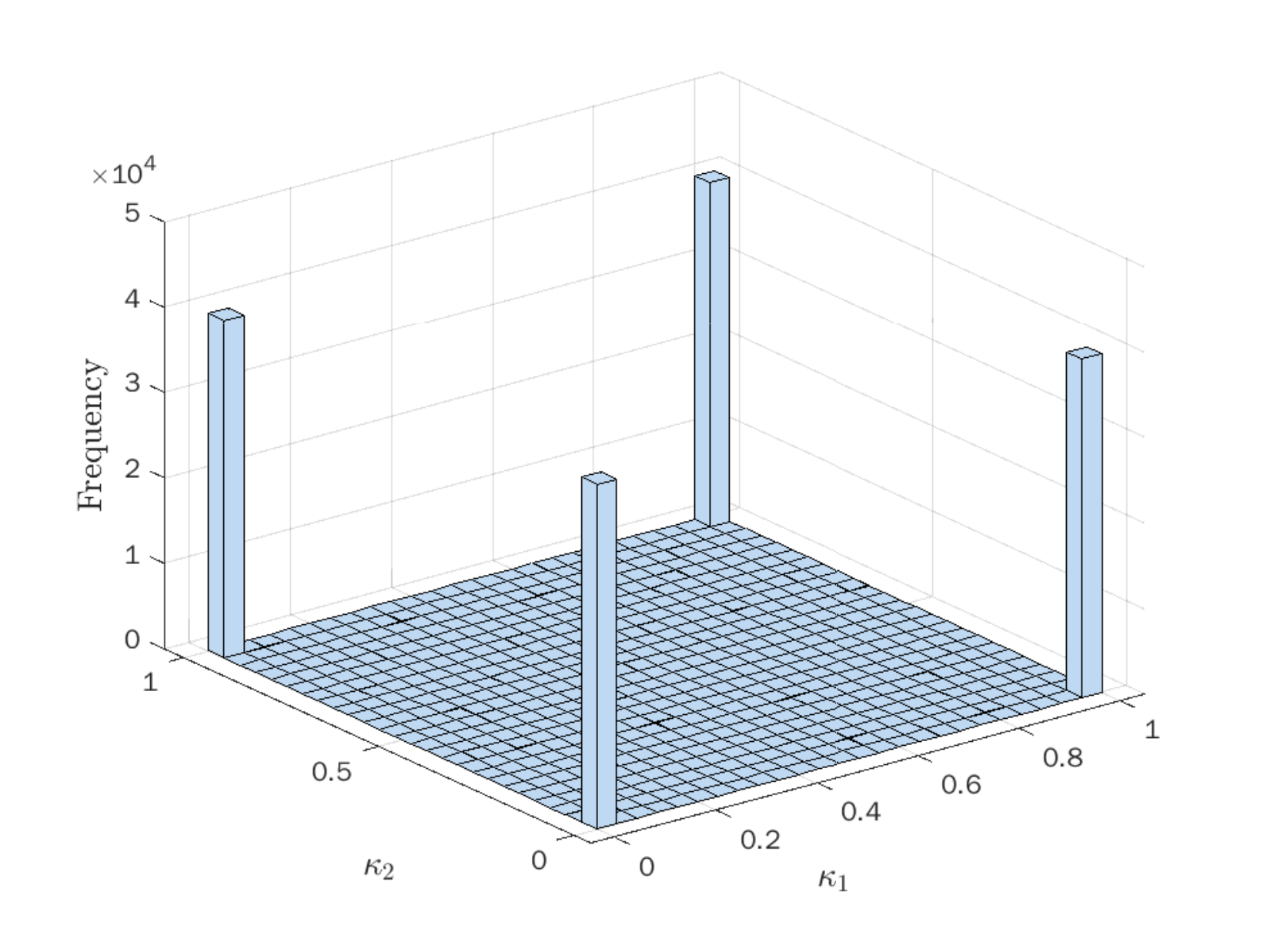}
		\caption{The distribution of the best response of the encoder $\gamma^{e}=(\kappa_1,\kappa_2)$ for given decoder actions. As it can be seen, $\kappa_1\kappa_2$ can only be any of $00$, $01$, $10$, and $11$.}
		\label{figure:kappaHist}
	\end{subfigure}
	\hfill
	\begin{subfigure}[t]{0.30\textwidth}
		\includegraphics[scale=0.35]{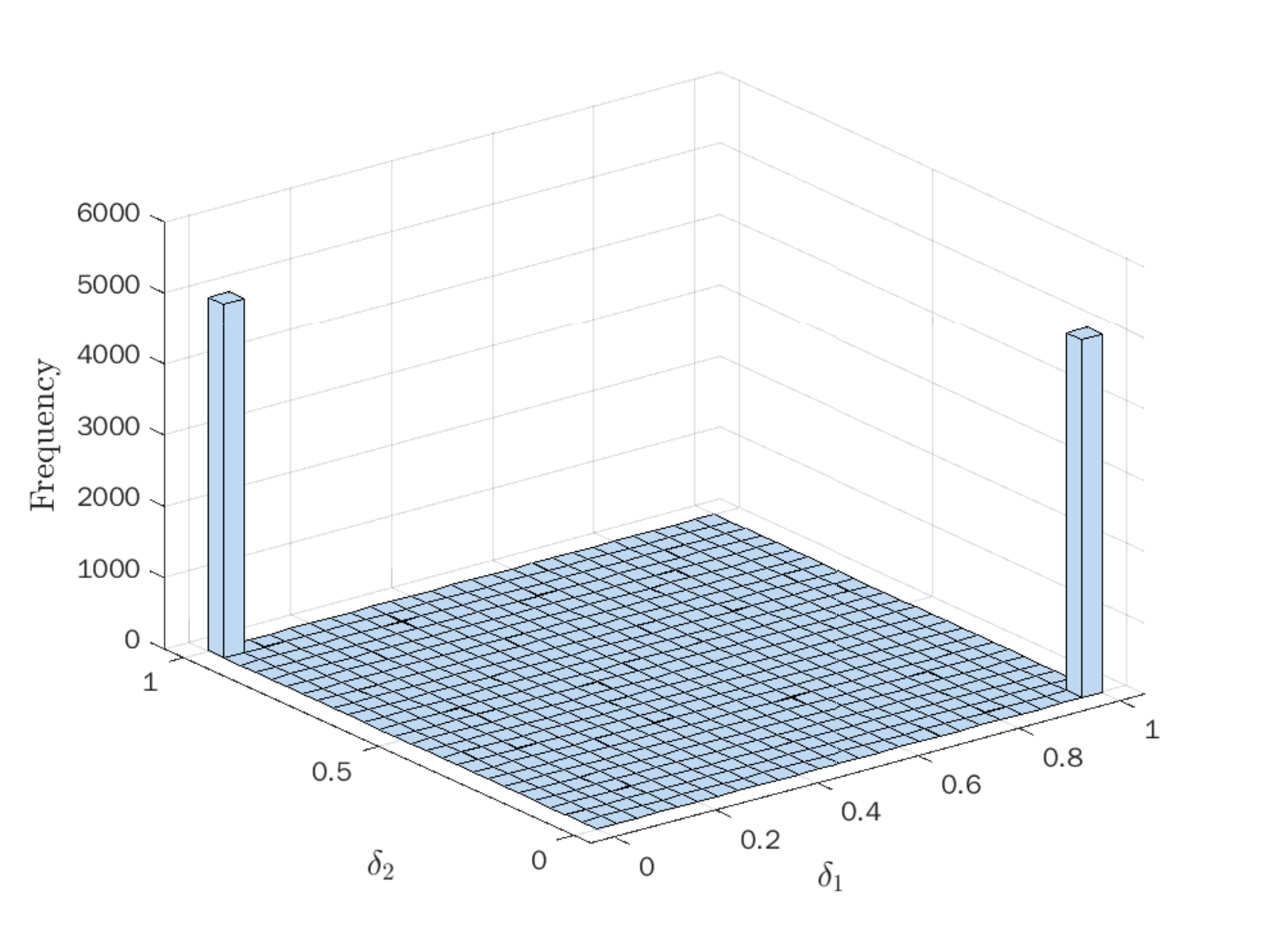}
		\caption{The distribution of the best response of the decoder $\gamma^{d_{\bf y}}=(\delta_1,\delta_2)$ for given encoder actions. As it can be seen, $\delta_1\delta_2$ is either $01$ or $10$.}
		\label{figure:deltaHist}
	\end{subfigure}
	\hfill
	\begin{subfigure}[t]{0.30\textwidth}
		\includegraphics[scale=0.63]{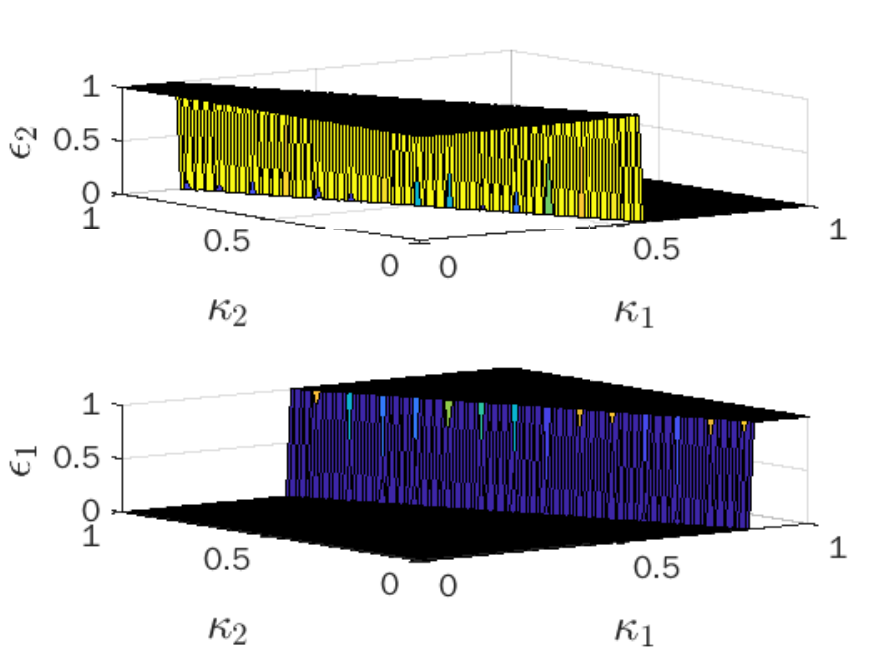}
		\caption{The best response of the decoder $\gamma^{d_{\bf x}}=(\epsilon_1,\epsilon_2)$ for given encoder actions. As it can be seen, $\epsilon_1\epsilon_2$ can only be any of $00$, $01$, and $10$.}
		\label{figure:epsilonSurf}
	\end{subfigure}
	\caption{Simulation results on the best responses of the players.}
	\label{fig:histogramExample}
\end{figure}

In this section, we validate our theoretical results via simulations. For all simulations, we let $a = 0.3$, $b = 0.1$, $c = 0.2$, and $d = 1-a-b-c=0.4$. We start by validating the best responses of the players as follows:
\begin{itemize}
    \item[(i)] First we let $\rho=1$. Then, for given decoder strategies $(\delta_1,\delta_2)\in[0,1]^2$ and $(\epsilon_1,\epsilon_2)\in[0,1]^2$, we calculate corresponding encoder costs by \eqref{reform_obj_enc} for every possible symmetric encoder actions $\gamma^{e}=(\kappa_1,\kappa_2)$, to find the optimal one, i.e., the maximizer. We repeat this process for every $\delta_1$, $\delta_2$, $\epsilon_1$, and $\epsilon_2$, which can take one of the $20$ evenly spaced values between $0$ and $1$. As illustrated in Fig. \ref{figure:kappaHist}, for $16\times10^4$ different combinations, $\kappa_1\kappa_2$ takes only four different values, which are $\delta_1\delta_2 = 00$, $\delta_1\delta_2 = 01$, $\delta_1\delta_2 = 10$, or $\delta_1\delta_2 = 11$. Thus, the best response of the encoder stated in Lemma~\ref{lemma:bestResponseEncoder} is proved numerically, too. Note that, in our simulations, when there are multiple optima, the encoder selects any of them randomly. 
    \item[(ii)] For a given encoder strategy $(\kappa_1,\kappa_2)\in[0,1]^2$, we calculate corresponding decoder costs by \eqref{reform_obj_dec} (indeed, only the mutual information $I({\bf y};\hat{\bf y})$ part) for every possible decoder actions $\gamma^{d_{\bf y}}=(\delta_1,\delta_2)$, to find the optimal one, i.e., the maximizer. We repeat this process for every $\kappa_1$ and $\kappa_2$, which can take one of the $100$ evenly spaced values between $0$ and $1$. As illustrated in Fig. \ref{figure:deltaHist}, for $10^4$ different $\kappa_1\kappa_2$ values, $\delta_1\delta_2$ takes only two different values, which are $\delta_1\delta_2 = 01$ or $\delta_1\delta_2 = 10$. Thus, the best response of the decoder stated in Lemma~\ref{lemma:bestResponseDecoder}.(i) is proved numerically, too. Note that, in our simulations, when there are multiple optima, the decoder selects any of them randomly. Thus, always optimal actions $\delta_1\delta_2 = 01$ or $\delta_1\delta_2 = 10$ are selected approximately equal number of times.
    \item[(iii)] Similar to the previous analysis, for a given encoder strategy $(\kappa_1,\kappa_2)\in[0,1]^2$, now we calculate corresponding decoder costs by \eqref{reform_obj_dec} (indeed, only the average distortion ${\bf E}\{d_H({\bf x},\hat{\bf  x})\}$ part) for every possible decoder actions $\gamma^{d_{\bf x}}=(\epsilon_1,\epsilon_2)$, to find the optimal one, i.e., the minimizer. We repeat this process for every $\kappa_1$ and $\kappa_2$, and obtain Fig. \ref{figure:epsilonSurf}. The optimal actions are $\epsilon_1\epsilon_2 = 00$, $\epsilon_1\epsilon_2 = 01$ or $\epsilon_1\epsilon_2 = 10$. Thus, the best response of the decoder stated in Lemma~\ref{lemma:bestResponseDecoder}.(ii) is proved numerically, too. Note that, by Table~\ref{table:optDecoderX}, $\epsilon_1\epsilon_2$ cannot be $11$ since $a+b\geq c+d$ is not satisfied for our selection.
\end{itemize}

\par
\begin{figure}
	\centering
	\begin{subfigure}[t]{0.43\textwidth}
		\includegraphics[scale=0.57]{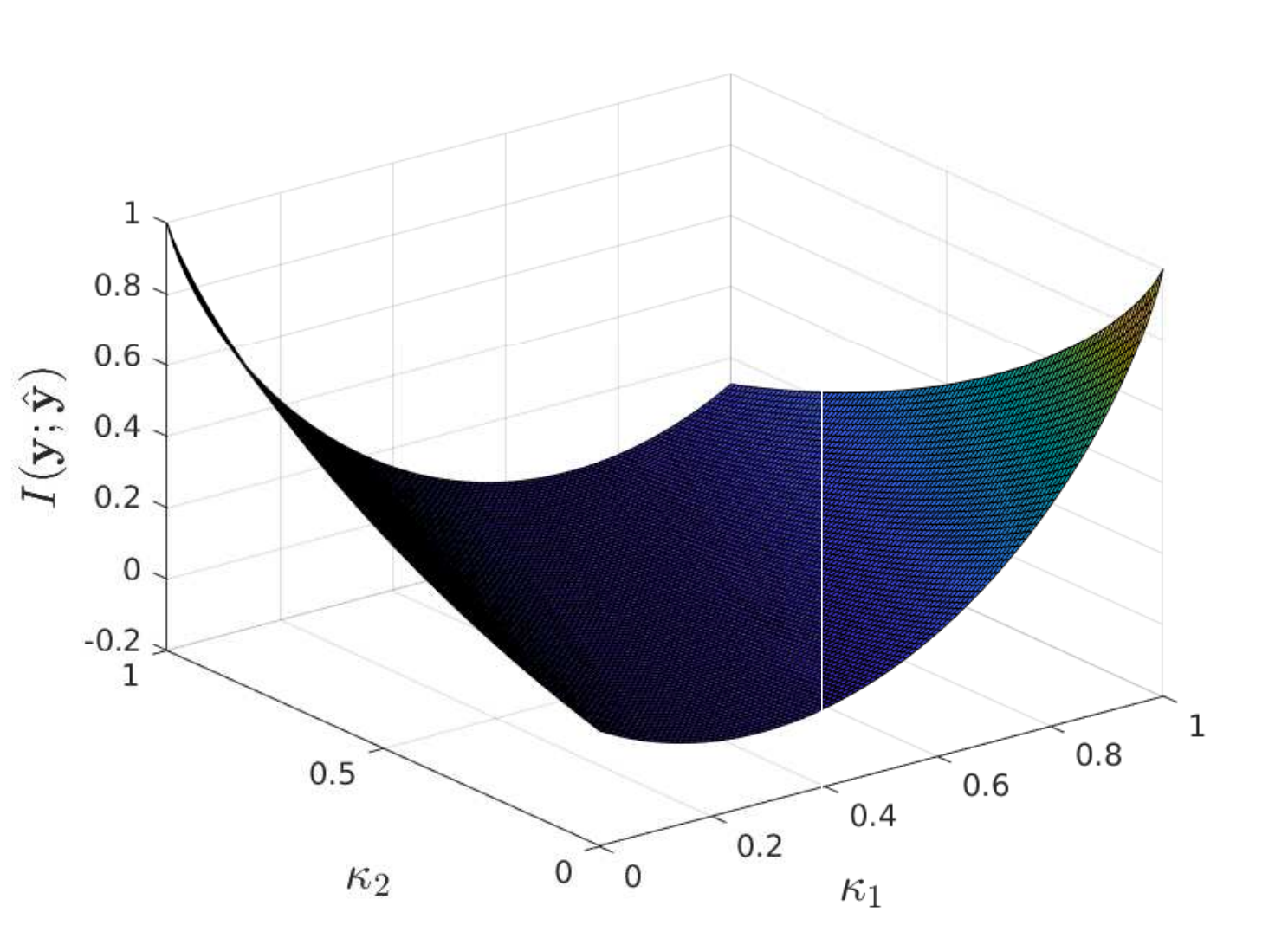}
		\caption{}
		\label{figure:mutualInfo}
	\end{subfigure}
	\hfill
	\begin{subfigure}[t]{0.43\textwidth}
		\includegraphics[scale=0.57]{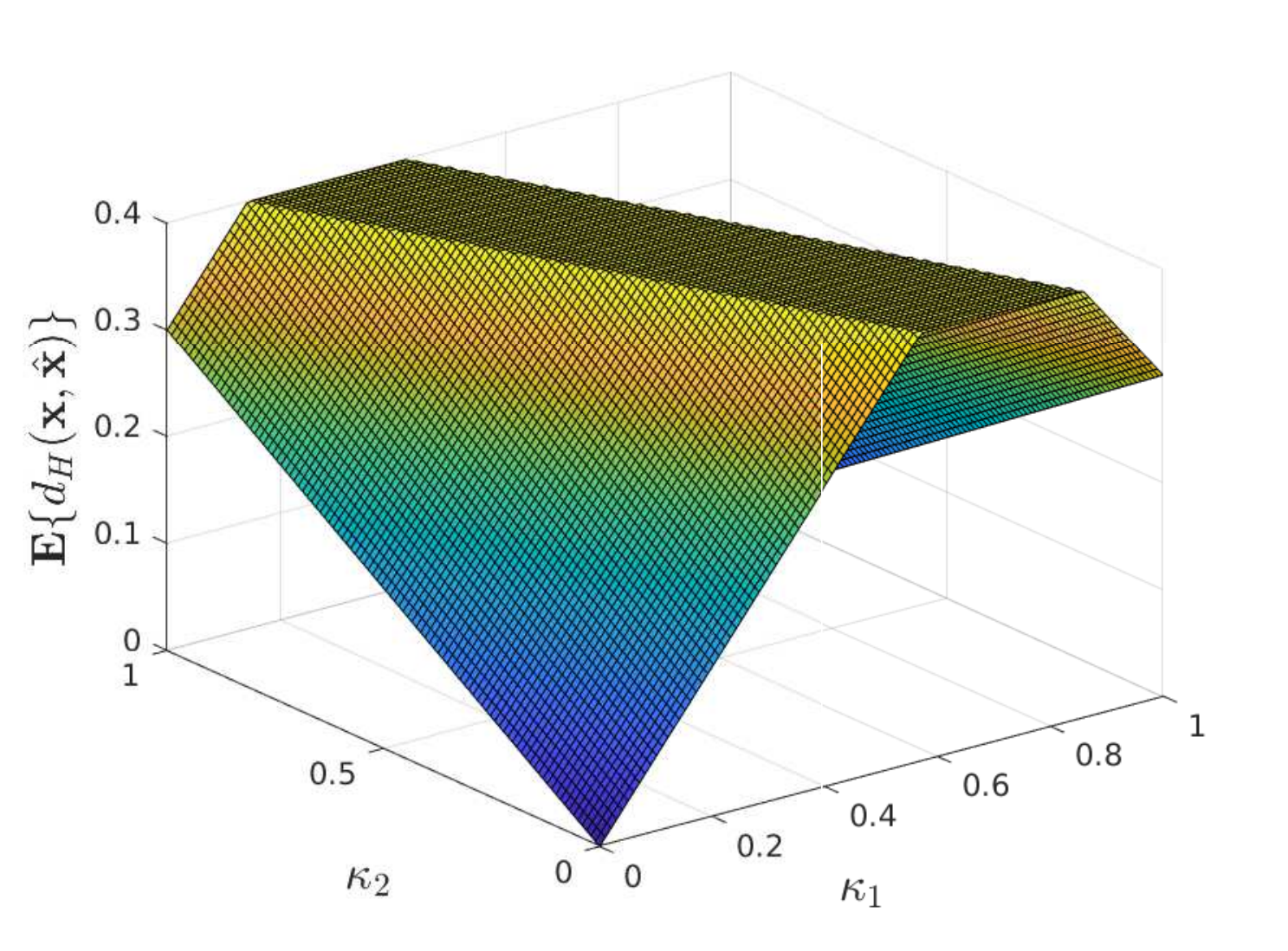}
		\caption{}
		\label{figure:hamming}
	\end{subfigure}
	\caption{$I({\bf y};\hat{\bf y})$ and ${\bf E}\{d_H({\bf x},\hat{\bf  x})\}$ as a function of the encoder strategy $(\kappa_1,\kappa_2)$ to analyze the Stackelberg equilibrium.}
	\label{fig:stackelbergExample}
\end{figure}

\begin{figure}
	\centering
	\begin{subfigure}[t]{0.43\textwidth}
		\includegraphics[scale=0.94]{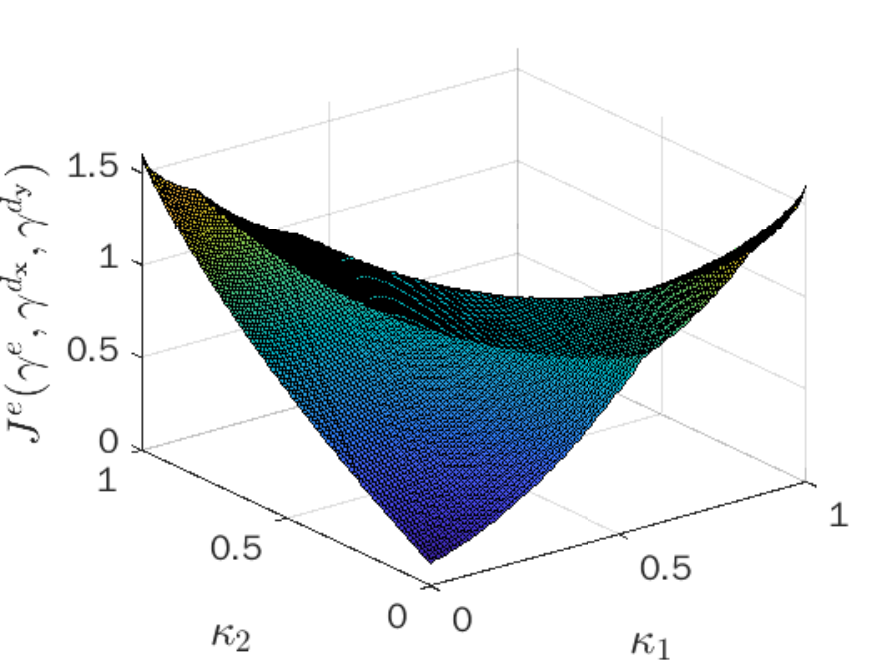}
		\caption{$\rho=2$.}
		\label{figure:rho2}
	\end{subfigure}
	\hfill
	\begin{subfigure}[t]{0.43\textwidth}
		\includegraphics[scale=0.94]{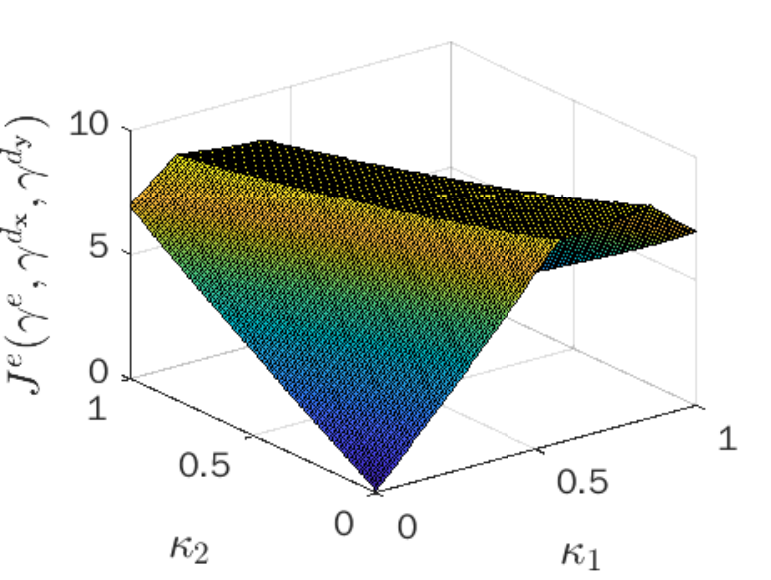}
		\caption{$\rho=20$.}
		\label{figure:rho20}
	\end{subfigure}
	\caption{$J^e(\gamma^e,\gamma^{d_{\bf x}},\gamma^{d_{\bf y}})$ as a function of the encoder strategy $(\kappa_1,\kappa_2)$ to analyze the Stackelberg equilibrium. As it can be seen, for small $\rho$, $\kappa_1\kappa_2 = 01$ and $\kappa_1\kappa_2 = 10$ are still optimal, whereas, the encoder selects intermediate $\kappa_1$ and $\kappa_2$ values as $\rho$ gets larger.}
	\label{fig:rhoExample}
\end{figure}

 After getting the best response maps of the players, we can utilize these results to obtain the Stackelberg equilibrium. 
In Fig. \ref{fig:stackelbergExample}, we plot $I({\bf y};\hat{\bf y})$ and ${\bf E}\{d_H({\bf x},\hat{\bf  x})\}$ as a function of the encoder strategy $(\kappa_1,\kappa_2)$ for given decoder strategies. In particular, Fig. \ref{figure:mutualInfo} illustrates the best response of the decoder-$\hat{\bf y}$ due to a Stackelberg assumption, i.e., $\delta_1\delta_2=01$ or $\delta_1\delta_2=10$ (via Lemma~\ref{lemma:bestResponseDecoder}), and the maximum $I({\bf y};\hat{\bf y})$ is achieved when $\kappa_1\kappa_2 = 01$ or $\kappa_1\kappa_2 = 10$ (see Theorem~\ref{thm:stackelberg}). In Fig. \ref{figure:hamming}, via Lemma~\ref{lemma:bestResponseDecoder}, the best response of the decoder-$\hat{\bf x}$ is considered due to a Stackelberg assumption. Since $\min\{a+b, c+d, a+d, b+c\}=b+c$, $\kappa_1\kappa_2 = 01$ or $\kappa_1\kappa_2 = 10$ is not in the optimal region to maximize ${\bf E}\{d_H({\bf x},\hat{\bf  x})\}$ (see Theorem~\ref{thm:stackelberg} and Remark~\ref{rem:techComm}). {The effect on this confusion can be observed for large value of $\rho$. Indeed, for small enough privacy weighting coefficient $\rho$, as it can be seen in Fig.~\ref{figure:rho2}, the maximizers of $I({\bf y};\hat{\bf y})$ are still the maximizers of $J^e(\gamma^e,\gamma^{d_{\bf x}},\gamma^{d_{\bf y}})$ in \eqref{reform_obj_enc}. On the other hand, for large enough privacy weighting coefficient $\rho$, ${\bf E}\{d_H({\bf x},\hat{\bf  x})\}$ gets more dominant in $J^e(\gamma^e,\gamma^{d_{\bf x}},\gamma^{d_{\bf y}})$ in \eqref{reform_obj_enc}. This case is illustrated in Fig.~\ref{figure:rho20}.}

\section{Conclusion and Future Research}

In this paper, we studied Nash and Stackelberg equilibria of privacy signaling games with binary alphabets {with single-bit transmission} between an encoder and a decoder with misaligned objectives. We {derived} the conditions under which Nash and/or Stackelberg equilibria exist.

Our model has several possible interesting extensions. {The most important question that needs to be answered is the extension of the framework beyond the single-bit transmission, that is to say, the transmitted messages are random vectors. Another interesting extension would be to consider scenarios with alternative objective functions and privacy criteria (e.g., log-loss function).} 





\appendices

\section{Proof of Lemma \ref{lemma:characterization}}\label{proof:lemma:characterization}
From the information structure of the stochastic encoder and decoder in \eqref{trans_prob_enc}-\eqref{trans_prob_dec_2} the constraint sets in \eqref{reform_obj_enc} and \eqref{reform_obj_dec} are clear. Hence it suffices to characterize only the mutual information and the loss function. First, we characterize the mutual information between ${\bf y}$ and $\hat{\bf y}$ which is given by
\begin{align}
I({\bf y};\hat{\bf y}) = \sum_{({\bf y},\hat{\bf y})\in\{0,1\}^2} \log\left({{\bf P}(\hat{y}|y) \over {\bf P}(\hat{y})}\right){\bf P}(\hat{y}|y){\bf P}(y).\label{mutual_information}
\end{align}
To do it, we need to compute
\begin{align}
&    {\bf P}(\hat{y}|y) = \sum_{({\bf z},{\bf x})\in\{0,1\}^2} {\bf P}(\hat{y},z,x|y) = \sum_{({\bf z},{\bf x})\in\{0,1\}^2} {\bf P}^{d_{\bf y}}(\hat{y}|z){\bf P}^{e}(z|x,y){\bf P}(x|y) ,\label{reprod_pmf}\\
&{\bf P}(\hat{y}) = \sum_{{\bf y}\in\{0,1\}} {\bf P}(\hat{y}|y){\bf P}(y),\label{marginal_output}\\
&{\bf P}({y}) = \sum_{{\bf x}\in\{0,1\}} {\bf P}(x,y),\label{marginal_public}\\
& {\bf P}(x|y)=\frac{{\bf P}(x,y)}{{\bf P}(y)}.\label{posterior_x}
\end{align}
Observe that given \eqref{joint_prob_input_messages}, then in \eqref{marginal_public} we obtain
\begin{align}
{\bf P}(y)=\begin{bmatrix}
q_1\\
1-q_1
\end{bmatrix},\label{comput_marginal_public}
\end{align}
where $q_1\triangleq{a+c}$. This in turn implies when substituted in \eqref{posterior_x} that
\begin{align}
{\bf P}(x|y)=\begin{bmatrix}
\frac{a}{q_1} & \frac{b}{1-q_1}\\
\frac{c}{q_1} & \frac{d}{1-q_1}
\end{bmatrix}.\label{comput_posterior_x}    
\end{align}
Substituting \eqref{comput_posterior_x} in \eqref{reprod_pmf}, and using \eqref{trans_prob_enc}, \eqref{trans_prob_dec_1} we obtain 
\begin{align}
{\bf P}(\hat{y}|y)=\begin{bmatrix}
\frac{at_1+ct_3}{q_1} & \frac{bt_2+dt_4}{1-q_1}\\
\frac{a(1-t_1)+c(1-t_3)}{q_1} & \frac{b(1-t_2)+d(1-t_4)}{1-q_1}
\end{bmatrix},\label{comput_reprod}
\end{align}
where ${t_i=\delta_1\kappa_i+\delta_2(1-\kappa_i)},~i=1,2,3,4$. Finally, using \eqref{comput_reprod} and \eqref{comput_marginal_public} in \eqref{marginal_output} we obtain
\begin{align}
{\bf P}(\hat{y})=\begin{bmatrix}
at_1+bt_2+ct_3+dt_4\\
1-(at_1+bt_2+ct_3+dt_4)
\end{bmatrix}.\label{comput_marginal_output}
\end{align}
The characterization of the mutual information in \eqref{char_mutual_info} is obtained by substituting \eqref{comput_marginal_public}, \eqref{comput_reprod}, \eqref{comput_marginal_output},  in \eqref{reprod_pmf}. 
Next we proceed to characterize the loss function (i.e., Hamming distance in \eqref{obj_fun_enc}, \eqref{obj_fun_dec}). Observe that
\begin{align}
   {\bf E}\{d_H({\bf x},\hat{\bf  x})\} &= \sum_{({\bf x},\hat{\bf  x})\in\{0,1\}^2} d_H(x,\hat{x}){\bf P}(x,\hat{x}) = \sum_{({\bf x},\hat{\bf  x},{\bf y},{\bf z})\in\{0,1\}^2} d_H(x,\hat{x}){\bf P}(x,\hat{x},y,z) \nn\\
   &= \sum_{({\bf x},\hat{\bf  x},{\bf y},{\bf z})\in\{0,1\}^2} d_H(x,\hat{x}) {\bf P}^{d_{\bf x}}(\hat{x}|z){\bf P}^{e}(z|x,y){\bf P}(x,y) = \eqref{eq:hammingLoss}.
\end{align}
This completes the derivation.

\section{Proof of Lemma \ref{lemma:bestResponseEncoder}}\label{proof:lemma:bestResponseEncoder}

Observe the following analysis on mutual information in ~\eqref{char_mutual_info} with respect to $P_1$ and $P_2$:
\begin{align}
    {\partial I({\bf y};\hat{\bf y}) \over \partial P_1} &= -\log(P_1+P_2) + \log(1-(P_1+P_2))  + \log(P_1) - \log(q_1-P_1) \,,\nn\\
    {\partial I({\bf y};\hat{\bf y}) \over \partial P_2} &= -\log(P_1+P_2) + \log(1-(P_1+P_2))  + \log(P_2) - \log(1-q_1-P_2) \,\nn.
\end{align}
Then, the Hessian of $I({\bf y};\hat{\bf y})$ with respect to $P_1$ and $P_2$ becomes
\begin{align}
    \Hessian\,I({\bf y};\hat{\bf y}) = {\partial^2 I({\bf y};\hat{\bf y}) \over \partial P_i \partial P_j} = {1\over\ln(2)}\begin{bmatrix}
-A+B & -A\\
-A & -A+C
\end{bmatrix}\,,
\end{align}
where $A\triangleq{1\over P_1+P_2}+{1\over 1-(P_1+P_2)}$, $B\triangleq{1\over P_1}+{1\over q_1-P_1}$, and $C\triangleq{1\over P_2}+{1\over 1-q_1-P_2}$. Note that, by the definitions in Lemma~\ref{lemma:characterization}, we have $t_i\in[0,1]$, which implies $P_1\leq q_1$ and $P_2 \leq 1-q_1$. Therefore, $B>A$ and $C>A$ hold, i.e., the Hessian matrix has positive diagonals. Furthermore, since $|\Hessian I({\bf y};\hat{\bf y})|>0$, Hessian matrix has two positive eigenvalues, i.e., it is a positive definite matrix. This proves the (joint) convexity of $I({\bf y};\hat{\bf y})$ with respect to $P_1$ and $P_2$.

Now consider this analysis with respect to $\kappa_1$ and $\kappa_2$. By using the definitions of the parameters, the Jacobian of $P_1$ and $P_2$ with respect to $\kappa_1$ and $\kappa_2$ can be calculated as
\begin{align}
    J \triangleq \begin{bmatrix}
{\partial P_1 \over \partial \kappa_1} & {\partial P_1 \over \partial \kappa_2}\\
{\partial P_2 \over \partial \kappa_1} & {\partial P_2 \over \partial \kappa_2}
\end{bmatrix} = (\delta_1-\delta_2)\begin{bmatrix}
 a & -c\\
-d & b
\end{bmatrix} \,,
\end{align}
and corresponding Hessian becomes
\begin{align}
    &\widetilde{\Hessian}\,I({\bf y};\hat{\bf y}) = {\partial^2 I({\bf y};\hat{\bf y}) \over \partial \kappa_i \partial \kappa_j} = J^T \Hessian\,I({\bf y};\hat{\bf y}) \, J \,.
\end{align}
Here, by utilizing the analysis of $\Hessian\,I({\bf y};\hat{\bf y})$, it can be similarly shown that $\widetilde{\Hessian}\,I({\bf y};\hat{\bf y})$ has also positive diagonals and determinant, thus a positive definite matrix, which shows the (joint) convexity of $I({\bf y};\hat{\bf y})$ with respect to $\kappa_1$ and $\kappa_2$.

Regarding the distortion part, it can be seen that the Hamming distortion in \eqref{eq:hammingLoss} is an affine function of $\kappa_1$ and $\kappa_2$. Consequently, $I({\bf y};\hat{\bf y})+\rho{\bf E}\{d({\bf x},\hat{\bf  x})\}$ is a (jointly) convex function of $\kappa_1$ and $\kappa_2$ since the first term is convex and the second term is affine in $\kappa_1$ and $\kappa_2$. 


As a result, for given decoder strategies $\gamma^{d_{\bf x}}$ and $\gamma^{d_{\bf y}}$, the encoder chooses its strategy $\gamma^{e}$ as one of its extreme points $\kappa_1\kappa_2=\{00,01,10,11\}$ which results in the maximum $J^e$. This completes the proof.

\section{Proof of Lemma \ref{lemma:bestResponseDecoder}}\label{proof:lemma:bestResponseDecoder}
Suppose that Assumption \ref{assumption:1} holds. Then we obtain the following.
\begin{itemize}
    \item[(i)] Using a similar approach in the proof of Lemma~\ref{lemma:bestResponseEncoder}, first the Jacobian of $P_1$ and $P_2$ with respect to $\delta_1$ and $\delta_2$, and then the Hessian of $I({\bf y};\hat{\bf y})$ with respect to $\delta_1$ and $\delta_2$ can be calculated, and it can be shown that $I({\bf y};\hat{\bf y})$ is a (jointly) convex function of $\delta_1$ and $\delta_2$.
Then, the decoder $\gamma^{d_{\bf y}}$ selects extreme points of $\delta_1$ and $\delta_2$ in order to maximize $I({\bf y};\hat{\bf y})$. In particular, to find the best response of the decoder $\gamma^{d_{\bf y}}$, it would suffice to compare the values of $I({\bf y};\hat{\bf y})$ at four different $\delta_1\delta_2$ pairs, i.e., $00$, $01$, $10$, and $11$, tabulated as follows.

\begin{table}[ht]
\centering
\begin{tabular}{|c|c|c|c|}
\hline
$\delta_1\delta_2$ & $t_i$ & $P_1$ & $P_2$ \\ \hline
$00$ & $0$ & $0$ & $0$ \\ \hline
$01$ & $1-{\kappa_i}$ & $a(1-\kappa_1)+c\kappa_2$ & $b(1-\kappa_2)+d\kappa_1$ \\ \hline
$10$ & ${\kappa_i}$ & $a\kappa_1+c(1-\kappa_2)$ & $b\kappa_2+d(1-\kappa_1)$ \\ \hline
$11$ & $0$ & $a+c$ & $b+d$ \\ \hline
\end{tabular}
\end{table}
From~\eqref{char_mutual_info}, it can be seen that $\delta_1\delta_2=00$ and $\delta_1\delta_2=11$ result in the same $I({\bf y};\hat{\bf y})$; and similarly, $\delta_1\delta_2=01$ and $\delta_1\delta_2=10$ result in the same $I({\bf y};\hat{\bf y})$. Note that, for $\delta_1\delta_2=00$, we obtain $I({\bf y};\hat{\bf y})=0$, and since the mutual information is a non-negative quantity, the best response of the decoder $\gamma^{d_{\bf y}}$ is $\delta_1\delta_2=01$ or $\delta_1\delta_2=10$.
    \item[(ii)] For a given encoder strategy ${\kappa_i}$, the decoder $\gamma^{d_{\bf x}}$ tries to minimize \eqref{eq:hammingLoss}. We can write \eqref{eq:hammingLoss} in a different way as follows: 
\begin{align}
{\bf E}\{d_H({\bf x},\hat{\bf  x})\} &= a (1-n_1) + b(1-n_2) + cn_3 + dn_4   \nn\\
&= a+b+\epsilon_1(-a\kappa_1-b\kappa_2+c\kappa_3+d\kappa_4) + \epsilon_2(-a(1-\kappa_1)-b(1-\kappa_2)+c(1-\kappa_3)+d(1-\kappa_4)) \nn\\
&= a+b+\epsilon_1(c+d-\kappa_1(a+d)-\kappa_2(b+c)) + \epsilon_2(-a-b+\kappa_1(a+d)+\kappa_2(b+c)).\nn
\end{align}
Thus, the optimal decoder strategy and corresponding distortion can be characterized as in Table~\ref{table:optDecoderX}. From there, it can be seen that the decoder chooses its strategy $\gamma^{d_{\bf x}}$ so that
\begin{align}
    {\bf E}\{d_H({\bf x},\hat{\bf  x})\} = \min\{a+b, c+d, \theta, 1-\theta\} \,.
\end{align}
\end{itemize}
This completes the proof.

\section{Proof of Theorem \ref{thm:stackelberg}}\label{proof:thm:stackelberg}

Suppose that Assumption \ref{assumption:1} holds. Then, for a given encoder action $\kappa_1\kappa_2$, the best response of the decoder $\gamma^{d_{\bf y}}$ is $\delta_1\delta_2=01$ (or equivalently $\delta_1\delta_2=10$). Then, the corresponding mutual information $I({\bf y};\hat{\bf y})$ becomes 
\begin{align}
I({\bf y};\hat{\bf y}) &= H_b (a+c) + H_b (a+b) + a\log(a) + b\log(b)+c\log(c)+d\log(d) \leq  H_b (a+c) , 
\end{align}
for $\kappa_1\kappa_2=00$ or $\kappa_1\kappa_2=11$, and 
\begin{align}
I({\bf y};\hat{\bf y}) &= H_b (a+c),
\end{align}
for $\kappa_1\kappa_2=01$ or $\kappa_1\kappa_2=10$. Thus, considering only the mutual information part, the optimal encoder strategy $\gamma^{e}$ is either $\kappa_1\kappa_2=01$ or $\kappa_1\kappa_2=10$.

Regarding the distortion part, the encoder aims to make it as great as possible, more precisely, the encoder tries to $\max_{\theta} \min\{a+b, c+d, \theta, 1-\theta\}$. Note that if $\min\{a+b, 1-(a+b), \theta, 1-\theta\}$ is $\theta$ or $1-\theta$, it is possible to obtain greater $\min\{a+b, c+d, \theta, 1-\theta\}$ by increasing or decreasing $\theta$, respectively. Thus, we obtain $\max_{\theta} \min\{a+b, c+d, \theta, 1-\theta\} = \min\{a+b, c+d\}$. Here, we can have two different conditions.

\begin{enumerate}
    \item[{\bf (C1)}] If $\min\{a+b, c+d, a+d, b+c\}$ is $a+b$ or $c+d$, then for $\kappa_1\kappa_2=01 \Rightarrow \theta=b+c$ or $\kappa_1\kappa_2=10 \Rightarrow \theta=a+d$, and we obtain ${\bf E}\{d_H({\bf x},\hat{\bf  x})\} = \min\{a+b, c+d, \theta, 1-\theta\}=\min\{a+b, c+d\}$, which is optimal for the encoder. Thus, both the mutual information part $I({\bf y};\hat{\bf y})$ and the distortion part ${\bf E}\{d_H({\bf x},\hat{\bf  x})\}$ are maximized when $\kappa_1\kappa_2=01$ or $\kappa_1\kappa_2=10$. Therefore, in this case, Stackelberg equilibrium strategies are given by \eqref{stack:strategies}.
    \item[{\bf (C2)}] If $\min\{a+b, c+d, a+d, b+c\}$ is $a+d$ or $b+c$, then for $\kappa_1\kappa_2=01 \Rightarrow \theta=b+c$ or $\kappa_1\kappa_2=10 \Rightarrow \theta=a+d$, and we get ${\bf E}\{d_H({\bf x},\hat{\bf  x})\} = \min\{a+b, c+d, \theta, 1-\theta\}=\min\{\theta, 1-\theta\}$, which is not an optimal choice for the encoder. The encoder selects its strategy so that either $a+b \leq \theta \leq c+d$ or $a+b \geq \theta \geq c+d$ hold depending on the prior. Therefore, even though $\kappa_1\kappa_2=01$ and $\kappa_1\kappa_2=10$ maximize $I({\bf y};\hat{\bf y})$, they are not optimal for maximizing ${\bf E}\{d_H({\bf x},\hat{\bf  x})\}$, thus the optimal encoder strategy which maximizes $J^e(\gamma^e,\gamma^{d_{\bf x}},\gamma^{d_{\bf y}})=I({\bf y};\hat{\bf y})+\rho{\bf E}\{d_H({\bf x},\hat{\bf  x})\}$ may not be these strategies. Even though for sufficiently small $\rho$ values, $\kappa_1\kappa_2=01$ and $\kappa_1\kappa_2=10$ characterizes the Stackelberg equilibrium, for large values of $\rho$, equilibrium strategies change. For sufficiently large $\rho$ values, the distortion (i.e., privacy) part becomes more dominant and the optimum encoder strategies are designed so that either $a+b \leq \theta \leq c+d$ or $a+b \geq \theta \geq c+d$ holds. In particular, for the privacy part, there are infinitely many equivalent encoder strategies $\kappa_1\kappa_2$ which satisfy either $a+b \leq \theta \leq c+d$ or $a+b \geq \theta \geq c+d$ resulting in the same distortion, where $\theta\triangleq \kappa_1(a+d)+\kappa_2(b+c)$ as defined earlier. Then, due to the convexity of mutual information in $\kappa_1\kappa_2$, the optimum encoder strategy lies at the boundary of the $\kappa_1\kappa_2$ region (which satisfy either $a+b \leq \theta \leq c+d$ or $a+b \geq \theta \geq c+d$).
\end{enumerate}
This completes the proof.

\section{Proof of Theorem \ref{thm:nash}}\label{proof:thm:nash}

Suppose that Assumption \ref{assumption:1} holds. Then, recall that the best responses of the players are as follows
\begin{itemize}
    \item for a given decoder strategy, the optimal encoder selects one of $\kappa_1\kappa_2=\{00,01,10,11\}$.
    \item for a given encoder strategy, the optimal decoder-$\hat{\bf y}$ selects either $\delta_1\delta_2 = 01$ or $\delta_1\delta_2 = 10$, which result in identical $I({\bf y};\hat{\bf y})$.
    \item for a given encoder strategy, the optimal decoder-$\hat{\bf x}$ selects one of $\epsilon_1\epsilon_2=\{00,01,10,11\}$.
\end{itemize}
Therefore, the game can be expressed in a normal form with $(\kappa_1\kappa_2,\epsilon_1\epsilon_2)$ (which correspond to the rows for the encoder strategies and the columns for the decoder-$\hat{\bf x}$ strategies, respectively) and corresponding costs\footnote{Note that comparing $J^e(\gamma^e,\gamma^{d_{\bf x}},\gamma^{d_{\bf y}})$ and $J^e(\gamma^e,\gamma^{d_{\bf x}},\gamma^{d_{\bf y}})-H_b (q_1)$ does not affect the equilibrium, and $M$ represents a non-positive value $M\triangleq H_b (a+b) + a\log(a) + b\log(b) + c\log(c)+d\log(d)$.} $J^e(\gamma^e,\gamma^{d_{\bf x}},\gamma^{d_{\bf y}})-H_b (q_1)$ and ${\bf E}\{d_H({\bf x},\hat{\bf  x})\}$.

\begin{table}[ht]
\centering
\begin{tabular}{|c|c|c|c|c|}
\hline
 & $00$ & $01$ & $10$ & $11$ \\ \hline
$00$ & $(M+\rho(a+b), a+b)$ & $(M, 0)$ & $(M+\rho, 1)$ & $(M+\rho(c+d), c+d)$\\ \hline
$01$ & $(\rho(a+b), a+b)$ & $(\rho(b+c), b+c)$ & $(\rho(a+d), a+d)$ & $(\rho(c+d), c+d)$ \\ \hline
$10$ & $(\rho(a+b), a+b)$ & $(\rho(a+d), a+d)$ & $(\rho(b+c), b+c)$ & $(\rho(c+d), c+d)$ \\ \hline
$11$ & $(M+\rho(a+b), a+b)$ & $(M+\rho, 1)$ & $(M, 0)$ & $(M+\rho(c+d), c+d)$ \\ \hline
\end{tabular}%
\end{table}

Here, it can be seen that,
\begin{itemize}
\item if $\min\{a+b, c+d, a+d, b+c\}=a+b$, then $(\kappa_1\kappa_2=01,\epsilon_1\epsilon_2=00)$ and $(\kappa_1\kappa_2=10,\epsilon_1\epsilon_2=00)$ are equilibrium points,
\item if $\min\{a+b, c+d, a+d, b+c\}=c+d$, then $(\kappa_1\kappa_2=01,\epsilon_1\epsilon_2=11)$ and $(\kappa_1\kappa_2=10,\epsilon_1\epsilon_2=11)$ are equilibrium points,
\item otherwise, a pure Nash equilibrium does not exist (excluding the trivial case of equal priors $a=b=c=d=0.25$ in which any strategy pair ends up an equilibrium).
\end{itemize}
This completes the proof.

\bibliographystyle{IEEEtran}
\bibliography{ecc2022Bibliography}

\begin{thebibliography}{10}
\providecommand{\url}[1]{#1}
\csname url@rmstyle\endcsname
\providecommand{\newblock}{\relax}
\providecommand{\bibinfo}[2]{#2}
\providecommand\BIBentrySTDinterwordspacing{\spaceskip=0pt\relax}
\providecommand\BIBentryALTinterwordstretchfactor{4}
\providecommand\BIBentryALTinterwordspacing{\spaceskip=\fontdimen2\font plus
\BIBentryALTinterwordstretchfactor\fontdimen3\font minus
  \fontdimen4\font\relax}
\providecommand\BIBforeignlanguage[2]{{%
\expandafter\ifx\csname l@#1\endcsname\relax
\typeout{** WARNING: IEEEtran.bst: No hyphenation pattern has been}%
\typeout{** loaded for the language `#1'. Using the pattern for}%
\typeout{** the default language instead.}%
\else
\language=\csname l@#1\endcsname
\fi
#2}}

\bibitem{mcdaniel-mclaughlin:2009}
P.~McDaniel and S.~McLaughlin, ``Security and privacy challenges in the smart
  grid,'' \emph{IEEE Security Privacy}, vol.~7, no.~3, pp. 75--77, 2009.

\bibitem{finster-baumgart:2015}
S.~Finster and I.~Baumgart, ``Privacy-aware smart metering: A survey,''
  \emph{IEEE Communications Surveys Tutorials}, vol.~17, no.~2, pp. 1088--1101,
  2015.

\bibitem{ny-2014}
J.~Le~Ny and G.~J. Pappas, ``Differentially private filtering,'' \emph{IEEE
  Transactions on Automatic Control}, vol.~59, no.~2, pp. 341--354, 2014.

\bibitem{gomez-vilardebo:2015}
J.~Gómez-Vilardebó and D.~Gündüz, ``Smart meter privacy for multiple users
  in the presence of an alternative energy source,'' \emph{IEEE Transactions on
  Information Forensics and Security}, vol.~10, no.~1, pp. 132--141, 2015.

\bibitem{zuxing:2019}
Z.~Li, T.~J. Oechtering, and D.~Gündüz, ``Privacy against a hypothesis
  testing adversary,'' \emph{IEEE Transactions on Information Forensics and
  Security}, vol.~14, no.~6, pp. 1567--1581, 2019.

\bibitem{nekouei:2019}
E.~Nekouei, T.~Tanaka, M.~Skoglund, and K.~H. Johansson,
  ``Information-theoretic approaches to privacy in estimation and control,''
  \emph{Annual Reviews in Control}, vol.~47, pp. 412--422, 2019.

\bibitem{lu-zhu:2020}
Y.~Lu and M.~Zhu, ``On privacy preserving data release of linear dynamic
  networks,'' \emph{Automatica}, vol. 115, p. 108839, 2020.

\bibitem{cavarec:2021}
B.~Cavarec, P.~A. Stavrou, M.~Bengtsson, and M.~Skoglund, ``Designing privacy
  filters for hidden {M}arkov processes,'' in \emph{European Control Conference
  (ECC)}, 2021.

\bibitem{SignalingGames}
V.~P. Crawford and J.~Sobel, ``Strategic information transmission,''
  \emph{Econometrica}, vol.~50, pp. 1431--1451, 1982.

\bibitem{misBehavingAgents}
I.~Shames, A.~M.~H. Teixeira, H.~Sandberg, and K.~H. Johansson, ``Agents
  misbehaving in a network: a vice or a virtue?'' \emph{IEEE Network}, vol.~26,
  no.~3, pp. 35--40, May 2012.

\bibitem{csLloydMax}
B.~Larrousse, O.~Beaude, and S.~Lasaulce, ``Crawford-{S}obel meet {L}loyd-{M}ax
  on the grid,'' in \emph{IEEE International Conference on Acoustics, Speech
  and Signal Processing (ICASSP)}, May 2014, pp. 6127--6131.

\bibitem{miklos2013value}
J.~Mikl{\'o}s-Thal and H.~Schumacher, ``The value of recommendations,''
  \emph{Games and Economic Behavior}, vol.~79, pp. 132--147, 2013.

\bibitem{recommSystemGame}
O.~Ben-Porat and M.~Tennenholtz, ``A game-theoretic approach to recommendation
  systems with strategic content providers,'' in \emph{International Conference
  on Neural Information Processing Systems (NeurIPS)}, 2018, p. 1118–1128.

\bibitem{signalSurvey}
J.~G. Riley, ``Silver signals: Twenty-five years of screening and signaling,''
  \emph{Journal of Economic Literature}, vol.~39, no.~2, pp. 432--478, June
  2001.

\bibitem{Sobel2009}
J.~Sobel, ``Signaling games,'' in \emph{Encyclopedia of Complexity and Systems
  Science}, R.~A. Meyers, Ed.\hskip 1em plus 0.5em minus 0.4em\relax Springer
  New York, 2009, pp. 8125--8139.

\bibitem{bayesianPersuasion}
E.~Kamenica and M.~Gentzkow, ``Bayesian persuasion,'' \emph{American Economic
  Review}, vol. 101, no.~6, pp. 2590--2615, Oct. 2011.

\bibitem{tacWorkSerkan}
S.~Sar{\i}ta\c{s}, S.~Y{\"{u}}ksel, and S.~Gezici, ``Quadratic
  multi-dimensional signaling games and affine equilibria,'' \emph{IEEE
  Transactions on Automatic Control}, vol.~62, no.~2, pp. 605--619, Feb. 2017.

\bibitem{CedricWork}
F.~Farokhi, A.~M.~H. Teixeira, and C.~Langbort, ``Estimation with strategic
  sensors,'' \emph{IEEE Transactions on Automatic Control}, vol.~62, no.~2, pp.
  724--739, Feb. 2017.

\bibitem{akyolITapproachGame}
E.~Akyol, C.~Langbort, and T.~Ba\c{s}ar, ``Information-theoretic approach to
  strategic communication as a hierarchical game,'' \emph{Proceedings of the
  IEEE}, vol. 105, no.~2, pp. 205--218, Feb. 2017.

\bibitem{omerHierarchial}
M.~O. Sayin, E.~Akyol, and T.~Ba\c{s}ar, ``Hierarchical multistage {G}aussian
  signaling games in noncooperative communication and control systems,''
  \emph{Automatica}, vol. 107, pp. 9--20, 2019.

\bibitem{dynamicGameSerkan}
S.~Sar{\i}ta\c{s}, S.~Y\"uksel, and S.~Gezici, ``Dynamic signaling games with
  quadratic criteria under {N}ash and {S}tackelberg equilibria,''
  \emph{Automatica}, vol. 115, p. 108883, May 2020.

\bibitem{strategicCommSideInfo}
M.~le~Treust and T.~Tomala, ``Strategic communication with decoder side
  information,'' in \emph{IEEE International Symposium on Information Theory
  (ISIT)}, 2021, pp. 2696--2701.

\bibitem{persuasionLimCap}
M.~L. Treust and T.~Tomala, ``Persuasion with limited communication capacity,''
  \emph{Journal of Economic Theory}, vol. 184, p. 104940, 2019.

\bibitem{serkanACC2020}
S.~{Sarıtaş}, G.~{Dán}, and H.~{Sandberg}, ``Passive fault-tolerant
  estimation under strategic adversarial bias,'' in \emph{American Control
  Conference (ACC)}, 2020, pp. 4644--4651.

\bibitem{ertanMismatch}
E.~Kaz{\i}kl{\i}, S.~Sar{\i}ta\c{s}, S.~Gezici, and S.~Y{\"{u}}ksel,
  ``Quadratic signaling with prior mismatch at an encoder and decoder:
  Equilibria, continuity and robustness properties,'' \emph{IEEE Transactions
  on Automatic Control}, pp. 1--1, 2022.

\bibitem{saritas-stavrou:2021}
S.~Sarıtaş, P.~A. Stavrou, R.~Thobaben, and M.~Skoglund, ``Quadratic
  signaling games with channel combining ratio,'' in \emph{IEEE International
  Symposium on Information Theory (ISIT)}, 2021, pp. 2690--2695.

\bibitem{farokhi:2015}
F.~Farokhi, H.~Sandberg, I.~Shames, and M.~Cantoni, ``Quadratic {G}aussian
  privacy games,'' in \emph{54th IEEE Conference on Decision and Control
  (CDC)}, 2015, pp. 4505--4510.

\bibitem{akyol:2015}
E.~Akyol, C.~Langbort, and T.~Başar, ``Privacy constrained information
  processing,'' in \emph{54th IEEE Conference on Decision and Control (CDC)},
  2015, pp. 4511--4516.

\bibitem{farokhi:2016}
F.~Farokhi and G.~Nair, ``Privacy-constrained communication,''
  \emph{IFAC-PapersOnLine}, vol.~49, no.~22, pp. 43--48, 2016.

\bibitem{kazikli:2020}
\BIBentryALTinterwordspacing
E.~Kazikli, S.~Gezici, and S.~Y{\"{u}}ksel, ``Quadratic privacy-signaling games
  and the {MMSE} information bottleneck problem for {G}aussian sources,''
  \emph{arxiv.org}, vol. abs/2005.05743v3, 2022. [Online]. Available:
  \url{https://arxiv.org/abs/2005.05743v3}
\BIBentrySTDinterwordspacing

\bibitem{privacyMetricSurvey}
I.~Wagner and D.~Eckhoff, ``Technical privacy metrics: A systematic survey,''
  \emph{ACM Comput. Surv.}, vol.~51, no.~3, June 2018.

\bibitem{cover-thomas:2006}
T.~M. Cover and J.~A. Thomas, \emph{Elements of Information Theory},
  2nd~ed.\hskip 1em plus 0.5em minus 0.4em\relax John Wiley \& Sons, Inc.,
  Hoboken, New Jersey, 2006.

\bibitem{basols99}
T.~Ba\c{s}ar and G.~J. Olsder, \emph{Dynamic Noncooperative Game Theory}.\hskip
  1em plus 0.5em minus 0.4em\relax Philadelphia, PA: SIAM Classics in Applied
  Mathematics, 1999.

\end{thebibliography}

\end{document}